\documentclass[preprint,showpacs,preprintnumbers,amsmath,amssymb,prb]{revtex4}
\usepackage{graphicx}% Include figure files
\usepackage{dcolumn}% Align table columns on decimal point
\usepackage{bm}% bold math
\usepackage{epsfig}
\usepackage{wasysym}
\usepackage{amsmath}
\begin{document}

\preprint{PREPRINT}

\title{Softness dependence of the Anomalies for the Continuous
  Shouldered Well potential}
\author{Pol Vilaseca and Giancarlo Franzese}
\affiliation{Departament de F\'{\i}sica Fonamental, Facultat de F\'{\i}sica,
Universitat de Barcelona, Diagonal 647, 08028, Barcelona, Spain.}
\email{pvilaseca@correu.ffn.ub.es, gfranzese@ub.edu}

\date{
%\today}
30  March 2010}

\begin{abstract}
By molecular dynamic simulations we study a system of particles interacting
through a continuous isotropic pairwise core-softened potential
consisting of a repulsive shoulder and an attractive well.
%[J. Mol. Liq. {\bf 136}, 267, (2007)].
The model displays a phase diagram with
three fluid phases, a gas-liquid critical point, a liquid-liquid
critical point, and anomalies in density, diffusion and structure. The
hierarchy of the anomalies is the same as for water. Here we study in
a systematic way the effect on the anomalies of varying the softness
of the potential. We find that, making the soft-core steeper and more penetrable, the
regions of density and diffusion anomalies contract in the $T-\rho$
plane, while the region of structural anomaly is weakly
affected. Therefore, 
a liquid can have anomalous structural behavior without having 
density or diffusion anomalies.
We show that, by considering
as effective distances those corresponding to 
the maxima of the first two peaks
of the radial distribution function $g(r)$ in the high-density liquid, we can generalize to
continuous two-scales potentials a criterion for
the occurrence of the anomalies of density and diffusion, originally
proposed for discontinuous potentials.
However, we observe that the knowledge of the structural behavior
within the first two coordination shells of the liquid is not enough
to establish, in general, the occurrence of the anomalies. By introducing the
density derivative of the the cumulative order integral of the excess entropy,
measuring shell by shell the amount of order in the liquid,
we show that  the
anomalous behavior is regulated by the structural order at
distances as large as the fourth coordination shell. By comparing the
results for different softness of the potential, we conclude that the
disappearing of the density and diffusion anomalies for the steeper
potentials is due to a more structured short-range order.
All these results increase our understanding on how, knowing the
interaction potential, we can 
evaluate the possible presence of anomalies for a liquid.

\end{abstract}


\maketitle

\section{Introduction}

Several liquids exhibit anomalous behaviours, i.e. behaviours that
differ from what is expected for normal liquids such as argon. Water
is the most common example of such systems. For example, while normal
liquids contract when they are cooled, water expands under
$T=4^{o}C$ at ambient pressure \cite{An76}.  

However, water is not unique in this respect.
The anomlaous behavior in density, in fact, has been observed also in
experimets for $Bi$ \cite{LosAlamos_r3}, $Ga$ \cite{LosAlamos_r3},
$Te$ \cite{Th76}, $S$ \cite{Sa67,Ke83}, $Ge_{15}Te_{85}$ \cite{Ts91}
and simulations for silica \cite{An00,Ru06b,Sh02,Po97}, silicon
\cite{Sa03} and $BeF_{2}$ \cite{An00}. All these
systems, including water, present a temperature of maximum density
(TMD) below which 
density decreases when temperature is lowered at constant pressure. 

Nevertheless, liquid water is anomalous also in other properties;
for example, in its dynamics and its structure.
While 
the diffuson coefficient $D$ of a normal liquid decreases as
density or pressure are increased, water is characterized by 
a region of the phase diagram where $D$ increases when the
pressure is increased.
Experiments show that the normal behaviour is restored at
pressures higher than $P\approx1.1$ kbar at $283K~$ \cite{An76}. 

Regarding the structure, normal liquids tend to become more
structurated when compressed. This can be quantified by 
two order parameters:
a translational order parameter $t$, quantifying the tendency of pairs
of molecules to adopt preferential separations, and an orientational
order parameter $Q_{l}$ 
that measures the tendency of a molecule and its 
nearest neigbhours to assume a specific
local arrangement, as considered by Steinhardt et al. \cite{St83}.
For a normal liquid $t$ and $Q_{l}$ increase with
pressure. Water,  instead, shows a region where the structural order parameters
 decrease for increasing pressure or density at constant $T$, i.e. the
 system becomes more disordered, as shown by  
Errington and Debendetti with molecular dynamics simualtions
\cite{Er01}.
Also simualtions for a model of silica display similar anomalies, as
shown by Shell et al. \cite{Sh02}. 

The sequence of the anomalies in water is well determined  in
experiments \cite{An76} and simulations \cite{Er01}.
For the $T-\rho$ phase diagram the water 
structural anomaly region is encompassing the diffusion anomaly region
which includes the density anomaly region. 

Despite the fact that these anomalies can be found also in other
liquids the sequence of anomalies may be 
different or may be the same. It is different,
for example, for silica where the anomalous difussion region
contains the structural anomaly regions that, in turn, includes
the density anomaly region \cite{Sh02}.  The sequence of anomalies is,
instead, the same for several isotropic models \cite{r15}.

Isotropic models are systems of
identical particles interacting through a central potential. 
It has been shown that isotropic potentials with a
repulsive soft-core, either a ramp or
a shoulder, have anomalies that resemble those of water
\cite{He70,St72,Ki96,special1,special2,special3,r32,r23,r25,r26,r27,r27b,r28,r28b,r28c,r29,r29b,r29c,r29d,r29e,r30,r30b,PhysRevE.71.031507,r31,PREGribova09,Fr01,r34,r35,r36}. 

%Potentials composed of a repulsive core with a region of softening,
%such as a ramp or a shoulder
%\cite{He70,St72,Ki96,special1,special2,special3,r23,r24,r25,r26,r27,r27b,r28,r28b,r28c,r29,r29b,r29c,r29d,r29e,Er03,r30b,r31,r32,Fr01,r34,r35,r36},
%are a simple aproach to understand the anomalous behaviours of
%flluids.  

Here we consider one of these potentials showing 
anomalies: the Continous Shoudered
Well potential (CSW), introduced in \cite{franzCSW1} and studied in
detail in \cite{r15}. This model can be considered as the continuous
version of the Discontinuous Shouldered Well potential (DSW) studied
in \cite{Fr01,r34,r35,r36}. The two potentials have been shown to
share the same qualitative phase diagram, with a liquid-iquid critical
point in the supercooled liquid phase. However, while the CSW has
anomalies, the DSW has not. 
The aim of the present study is to understand the origin of this difference.
In particular, we tune the CSW potential in such a way to approximate,
at least around the repulsive shoulder, the DSW, changing in this way 
the softness of the potential. Our results show that the continuous
tuning of the CSW induces a continuous change on the regions of
density and diffusion anomaly, but not a relevant effect on the
structural anomaly region. When the CSW approches the limit of the
DSW, the  regions of density and diffusion anomaly shrink as they
would disappear. 

The paper is organized as follows. We present the model 
in Sec. 2 and give details about the simulations in Sec 3. 
In Sec 4 we present our results about the phase diagram, 
and in Sec 5 our results about the anomalies, first with
a criteria based on the excess entropy and then with a direct
calculation. In Sec 6 we discuss the results, we generalize some
proposed criteria and we study the effect of the long-range order. In
Sec 7 we give our conclusions.  

\section{The Model}

\begin{figure}%[ht]
\begin{center}\includegraphics[clip=true,scale=0.375]{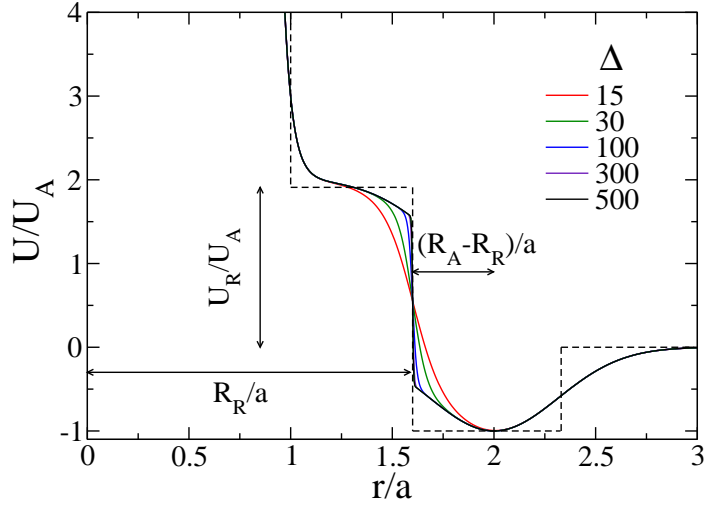}
\end{center}
\caption{Continuous lines represent the CSW potential for different
  values of the parameter $\Delta$ (slope at $r=R_{R}$). Dotted black
  line represents the discountinuous shouldered well potential (DSW)
  \cite{Fr01,r34,r35,r36}. By increasing the values of $\Delta$ the
  CSW potential tends to DSW around $R_{R}$.} 
\label{f1}
\end{figure}

We study the CSW model \cite{franzCSW1} consisting of a set of
identical particles interacting through the isotropic pairwise
potential given by 

%\begin{equation*}
% U(r)=\frac{U_{R}}{1+\exp\left(\Delta(r-R_{R})/a\right)}\\-U_{A}\times
%\end{equation*}
%\begin{equation}
% \exp\left[-\frac{(r-R_{A})^{2}}{2\delta_{A}^{2}}\right]+\left(\frac{a}{r}\right)^{24}\,
%\end{equation}
%
\begin{equation}
 U(r)=\frac{U_{R}}{1+\exp\left(\Delta(r-R_{R})/a\right)}-U_{A}
 \exp\left[-\frac{(r-R_{A})^{2}}{2\delta_{A}^{2}}\right]+\left(\frac{a}{r}\right)^{24}\,
\end{equation}
where $a$ is the diameter of the particles, $R_{A}$ and $R_{R}$ are
the distance of the attractive minimum and the repulsive
radius, respectivelly, $U_{A}$ and $U_{R}$ are 
the energies of the
attractive well and the repulsive shoulder, respectivelly, $\delta_{A}^{2}$ is the
variance of the Gaussian centered in $R_{A}$, and $\Delta$ is the
parameter which controls the slope between the shoulder and the well
at $R_{R}$ (Fig.~\ref{f1}). 
Varying the parameters the potential can
be tuned from a repulsive shoulder to a deep double well. 
In particular, by increasing $\Delta$ the soft-core repulsion becomes
more penetrable near the minimum of the attractive well, and the
softness of the potential increases for $r>R_R$ and decreases for $r<R_R$.

The set of
 values chosen for our analysis are as in \cite{franzCSW1,r15},
 $U_{R}/U_{A}=2$, $R_{R}/a=1.6$, $R_{A}/a=2$, 
 $\left(\delta_{A}/a\right)^{2}=0.1$, while we tune $\Delta$.
In order to reduce the
computational effort, we establish a cuttoff for $U(r)$ at
$r_{c}/a=3$, because $U(r)\rightarrow0$ for $r/a>3$
for all the considered values of $\Delta$.
%We also consider the related discontinuous shouldered well potential
%(DSW) (Fig.~\ref{f1})\cite{Fr01,r34,r35,r36}, that has been shown to
%display the same phase diagram as for the CSW. The DS, however, does
%not show density anomally. To understand why the density anomally is
%present in the CSW but absent in the DSW, we tune the CSW potential in
%such a way to approximate locally the DSW. 
In particular, we 
%chose to
%vary the parameter $\Delta\equiv\Delta/a$, determining the slope
%of the repulsive shoulder at $R_{R}$. We 
consider the values
$\Delta=15,30,100,300,500$ (Fig.~\ref{f1}) going from the case
$\Delta=15$ studied in Ref.~\cite{franzCSW1},\cite{r15} to slopes
that aproach the infinite value of DSW. 

For $\Delta=15$ this model is known to present anomalies in density, diffusion and
 structure \cite{franzCSW1,r15}. Its phase diagram displays a
 liquid-gas phase transition and a liquid-liquid phase transition,
 both ending in critical points \cite{franzCSW1,r15}.

\section{Molecular Dynamics}

For every chosen value of $\Delta$ we simulate the behaviour of
the system governed by the CSW potential using standard molecular
dynamics techniques \cite{Frenkel}. We perform simulations in the
\textit{NVT} ensemble for $N=1372$ particles interacting in a cubic
box of volume $V$ with periodic boundary conditions at temperature
$T$. We use 
%a simple 
the Allen thermostat keeping $T$ constant by
reescaling the velocities of the particles at each time step by a
factor $(T/\mathcal{T})^{\frac{1}{2}}$, where $\mathcal{T}$ is the
instantaneous kinetic temperature and $T$ the fixed temperature of the
thermal bath \cite{r38}. We integrate the equations of motion using
velocity Verlet integrator \cite{r38}. By fixing $\rho$ and $T$, we
calculate the pressure in terms of the second virial coefficient
\cite{r38}. 
All calculations are presented in dimensionless reduced units in terms
of the particle diameter and the depth of the attractive well:
$P^{*}\equiv Pa^{3}/U_{A}$, $T^{*}\equiv k_{B}T/U_{A}$,
$\rho^{*}\equiv \rho a^{3}$, and $D^{*}\equiv D(m/a^{2}U_{A})^{1/2}$,
where $D$ is the diffusion coefficient.

We use a simulation time step $\delta t^{*}=3.2\times10^{-3}$ defined in units
of $(a^{2}m/U_{A})^{1/2}$ (of the order of $\approx1.7\times10^{-12}
s$ for water-like molecules and $\approx2.1\times10^{-12} s$ for
argon-like atoms). 
Simulations are equilibrated after $10^{4}$ time steps, followed by
$10^{6}$ time steps during which thermodynamic quantities are computed
every 100 time steps. Positions and velocities are stored every 1000
time steps for the calculations of structure and dynamic quantities. 
For each value of $\rho^{*}$ and $T^{*}$ we average over two runs,
starting from independent configurations. 

Near the liquid-liquid critical point (described in the next section) we observe that
the liquid phase is metastable with respect to the crystal phase. The
nucleation of the crystal is marked by a subsequent drop in the
potential energey \cite{r34}. The lifetime of the metastable phase is
defined as the time between the end of the equilibration and the
energy drop.

\section{The phase diagram}

\begin{figure}%[ht]
\begin{center}\includegraphics[clip=true,scale=0.375]{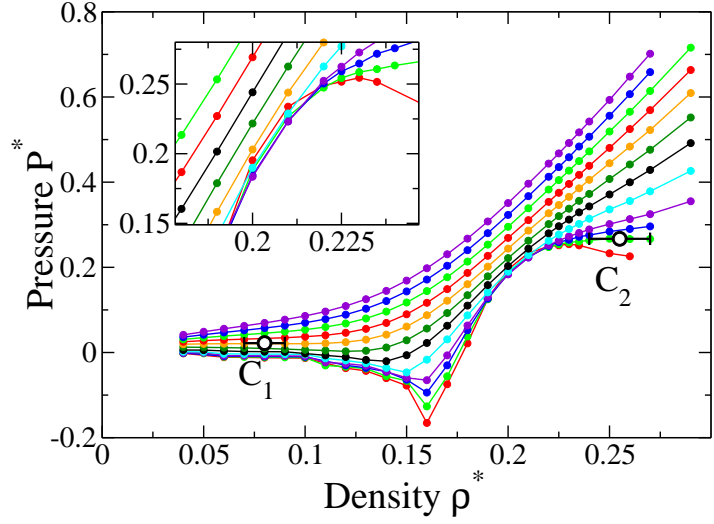}
\end{center}
\caption{Isotherms in the $P^{*}-\rho^{*}$ phase diagram from
  simulations for CSW potential with $\Delta=30$ 
(from top to  bottom, $T^{*}=1.4$, $1.3$, $1.2$, $1.1$, $1.0$,
  $0.9$, $0.8$, $0.7$, $0.6$, $0.55$, $0.5$ and $0.45$).
At low $T^{*}$, isotherms have no monothonic behaviour and show two van
  der Waals loops, corresponing to a liquid-gas (low densities) and a
  liquid-liquid (higher densities) first order phase transitions both
  ending in critical points ($C_1$ and $C_2$ respectivelly). 
 Inset: the
  crossing of the isotherms signals the presence of the density
  anomally. In this and the following figures errors, where not shown,
  are smaller than symbol size.} 
\label{f2}     
\end{figure}

First we verify that our simulations reproduce the known phase diagram
for the model with $\Delta=15$ \cite{franzCSW1,r15}.  
Next, we calculate the phase diagram for the other considered values
of $\Delta$, finding always the same qualitative behaviour
(Fig.\ref{f2}). In particular, we find two regions with
coexisting fluid phases: (i) gas and liquid at low $\rho^{*}$, (ii)
low density liquid (LDL) and high density liquid (HDL) at higher
$\rho^{*}$. 
The two coexistence regions end in critical points:  $C_{1}$ for
gas-liquid coexistence,  $C_{2}$ for LDL-HDL coexistence (Table
\ref{t1}). 
For $\Delta<100$ the 
HDL phase is metastable with respect to the crystal,
but with a lifetime long enough to allow us to equilibrate the liquid around $C_{2}$.
For $\Delta\geq100$ the HDL lifetime is too short to equilibrate the 
liquid around 
$C_{2}$ and we extrapolate the location of $C_{2}$ from higher
$T^{*}$ isotherms. 

 \begin{table*}[htp!]
\centering
\begin{tabular}{|c||c|c|c|c|c|c|}
\hline
  & $T^{*}_{C_1}$ & $P^{*}_{C_1}$ & $\rho^{*}_{C_1}$ & $T^{*}_{C_2}$ & $P^{*}_{C_2}$  & $\rho^{*}_{C_2}$\\ \hline \hline
$\Delta=15$ & $0.95\pm0.06$ & $0.019\pm0.008$ & $0.08\pm0.03$ & $0.49\pm0.01$ & $0.285\pm0.007$ & $0.247\pm0.008$ \\ 
$\Delta=30$ & $1.01\pm0.07$ & $0.022\pm0.008$ & $0.08\pm0.03$ & $0.50\pm0.01$ & $0.267\pm0.005$ & $0.255\pm0.009$ \\ 
$\Delta=100$ & $1.06\pm0.04$ & $0.025\pm0.005$ & $0.08\pm0.03$ & $0.53\pm0.02$ & $0.243\pm0.005$ & $0.262\pm0.008$ \\ 
$\Delta=300$ & $1.06\pm0.05$ & $0.027\pm0.009$ & $0.09\pm0.02$ & $0.51\pm0.01$ & $0.231\pm0.007$ & $0.271\pm0.008$ \\ 
$\Delta=500$ & $1.08\pm0.06$ & $0.027\pm0.008$ & $0.09\pm0.03$ & $0.52\pm0.01$ & $0.204\pm0.007$ & $0.272\pm0.008$ \\ 
DSW & $1.24\pm0.01$ & $0.030\pm0.001$ & $0.09\pm0.02$ & $0.69\pm0.02$ & $0.110\pm0.002$ & $0.280\pm0.020$ \\ 
\hline
\end{tabular}
 \caption{Temperatures $T^{*}_{C_1}$ and $T^{*}_{C_2}$, pressures
   $P^{*}_{C_1}$ and $P^{*}_{C_2}$, densities $\rho^{*}_{C_1}$ and
   $\rho^{*}_{C_2}$ (in reduced units) corresponding to the gas-liquid
   critical
   points $C_{1}$ and the LDL-HDL critical point $C_{2}$, respectivelly, for different
   values of $\Delta$ for the CSW potential. 
We
estimate the critical parameters 
%$T^{*}_{C_{i}}$, $p^{*}_{C_{i}}$ and
%$\rho^{*}_{C_{i}}$ of $C_{i}$, for $i=1,2$, 
by using the condition
$\left(\partial^{2}P/\partial\rho^{2}\right)|_T=0$
at $C_{1}$ and $C_{2}$. 
We confirm the data for $\Delta=15$ from Ref.~\cite{franzCSW1,r15}.
For sake of
   comparison, we indicate also the values for 
the DSW potential with 
parameters $U_{R}/U_{A}=2$, 
$R_{R}/a=1.6$,  $R_{A}/a=2$, and $(\delta_{A}/a)^2=0.1$ from
Ref.~\cite{Fr01,r34,r35,r36}}.   
\label{t1}
\end{table*} 

As  $\Delta$ is increased both $C_{1}$ and $C_{2}$ tend to the
corresponding values for the DSW potential. This result is consistent
with the idea that the DSW can be seen as a limiting case of the
familly of CSW potentials studied here. However, temperature and
pressure of the LDL-HDL critical point for $\Delta=500$ are still far from
the values for the DSW potential. 
 
\section{Hierarchy of the anomalies for different $\Delta$}

\begin{figure}%[ht]
\begin{center}\includegraphics[clip=true,scale=0.375]{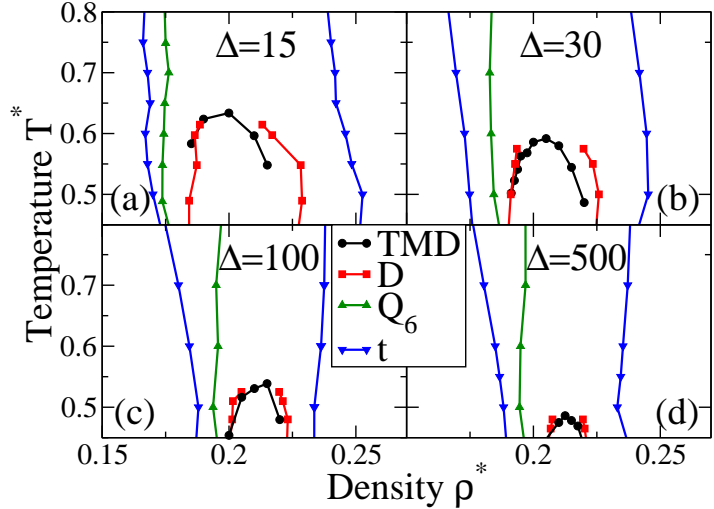}
\end{center}
	\caption{Hierarchy of anomalies (as defined in
          Sec.~V) in the $T^{*}-\rho^{*}$ plane,
          plotted for (a)  $\Delta=15$, (b) $\Delta=30$, (c)
          $\Delta=100$ and (d) $\Delta=500$. As the value of $\Delta$
          increases, both the regions of density anomaly ($\bullet$) and
          diffusion anomaly ($\blacksquare$) contract with the diffusion
          anomaly region always encompassing the TMD line. The region
          of structural anomaly (between $\blacktriangle$-symbols for
          $Q_6$ maxima and the high-$\rho^*$
          $\blacktriangledown$-symbols for $t$ minima) is only weakly affected.} 
\label{f7}       
\end{figure}

Barros de Oliveira et al. \cite{r15} show that the CSW
potential has anomalies in density, diffusion and
structure. Furthermore, they found a 
hierarchy of the anomalies as reported 
for other two-scale potentials and for the
SPC/E water. Here we study in a systematic way how these anomalies
depend on the parameter $\Delta$
of the CSW potential. In Fig.~\ref{f7} we summarize all our results. In the
next subsections we discuss the details of Fig.~\ref{f7}.

\subsection{Excess entropy}

As a starting point for the exploration of the anomalous regions of
the potential and in order to predict whether a given anomaly will
appear, we perform an analysis of the excess entropy.
The excess entropy $s^{\rm ex}$ is defined as the difference between
the entropy $s$ of a real fluid at a given $T$ and $\rho$ and the
entropy of an ideal gas $s^{\rm ig}$ under the same conditions,
$s^{\rm ex}\equiv s-s^{\rm ig}$, and quantifies the entropy lost due to
interactions and correlations between the particles of the fluid. 
It has been shown that the excess entropy allows an approximate
estimate of the regions where density, diffusion and
structure are anomalous.
In particular, Errington et al. \cite{errington:244502} proposed that the condition
under which the density anomaly appears 
is given by $\Sigma_{\rm ex}>1$, where by definition is  
$\Sigma_{\rm
  ex}\equiv(\partial s^{\rm ex}/\partial \ln\rho)_{T}$.
The diffusion anomaly can
also be predicted, following Rosenfeld's empirical parametrization
\cite{Ro99}, by the condition $\Sigma_{\rm ex}>0.42$. Structural
anomaly is determined by the criterion $\Sigma_{\rm ex}>0$, because
for normal fluids the 
excess entropy is always decreasing when density increases
isothermically \cite{Ra71}.

\begin{figure}%[ht]
\begin{center}\includegraphics[clip=true,scale=0.375]{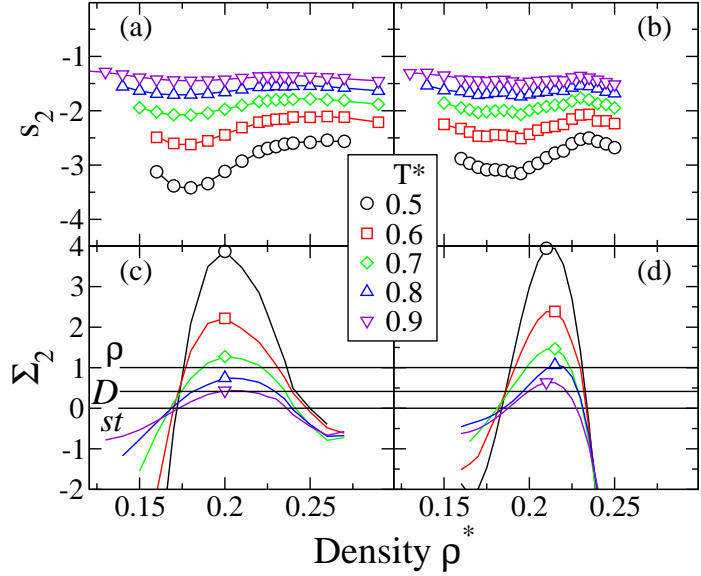}
\end{center}
		\caption{Excess entropy $s_{2}$ (upper panels) and its derivative
                  $\Sigma_{2}$ (lower panels)
%with respect to the logarithm of
%                 density 
for the potentials corresponding to
                  $\Delta=30$ (left panels) and $\Delta=500$ (right
                  panels), for $T^*=0.5$ ($\ocircle$), 0.6 ($\Box$), 0.7
                  ($\lozenge$), 0.8 ($\vartriangle$), 0.9
                  ($\triangledown$). 
Horizontal solid lines in the bottom panels
                  are the thresholds for $\Sigma_{2}$ for the anomalies
                  (from top to bottom for density, diffusion and
                  structural anomaly, respectivelly). As described in the text,
                  these criteria overestimate the regions of
                  anomalies.} 
\label{f8}             
\end{figure}

The evaluation of the excess entropy migth be computationally
expensive. However, it can be aproximated directly by the two-body
contribution term  
\begin{equation}
 s_{2}\equiv-2\pi\rho\int[g(r)\ln g(r)-g(r)+1]r^{2}dr
\label{2}
\end{equation}
which depends only on the radial distribution function and represents
the dominant contribution to the excess entropy
\cite{Ro99,Gr52,Ra71,Ba89} (between $85\%$ and $95\%$ of $s^{\rm ex}$
for a Lennard-Jones fluid \cite{Ba89,Cha06}). 
Since the integrand of Eq.(\ref{2})
goes to zero for large $r$, i.e. where $g(r)$ goes to
$1$, we calculate $s_2$ up to a cutoff given by half the
simulation-box size, distance at which we check that  $g(r)\approx 1$ 
within the range of $T$ and $\rho$ considered in this work. 
From Eq.(\ref{2}) we define and calculate
$\Sigma_{2}\equiv(\partial s_{2}/\partial \ln\rho)_{T}$.
We observe that the predicted regions of anomalies get narrow as the
slope $\Delta$ increases (Fig.~\ref{f8}). 

Since $s_{2}<s^{\rm ex}$, then it is always $\Sigma_{\rm ex}<\Sigma_{2}$.
%~\cite{errington:244502}.
Therefore, the criteria $\Sigma_{2}>1$, $\Sigma_{2}>0.42$ and
$\Sigma_{2}>0$ overestimate the regions of density, diffusion and
structural anomalies, respectivelly, with about $30\%$ uncertainty,
according to Rosenfeld \cite{Ro99}.  As we will discuss in Section VI, 
we
observe that the anomalous regions predicted in Fig.~\ref{f8} are
indeed an overstimate of the one we find as a
result of the detailed analysis presented in the next
sections. Nevertherless, the approximate criteria based on
$\Sigma_{2}$ give a preliminary idea of where the anomaly occur.

%\subsection{Molecular Dynamics}

\subsection{The density anomaly}

\begin{figure}%[ht]
\begin{center}\includegraphics[clip=true,scale=0.65]{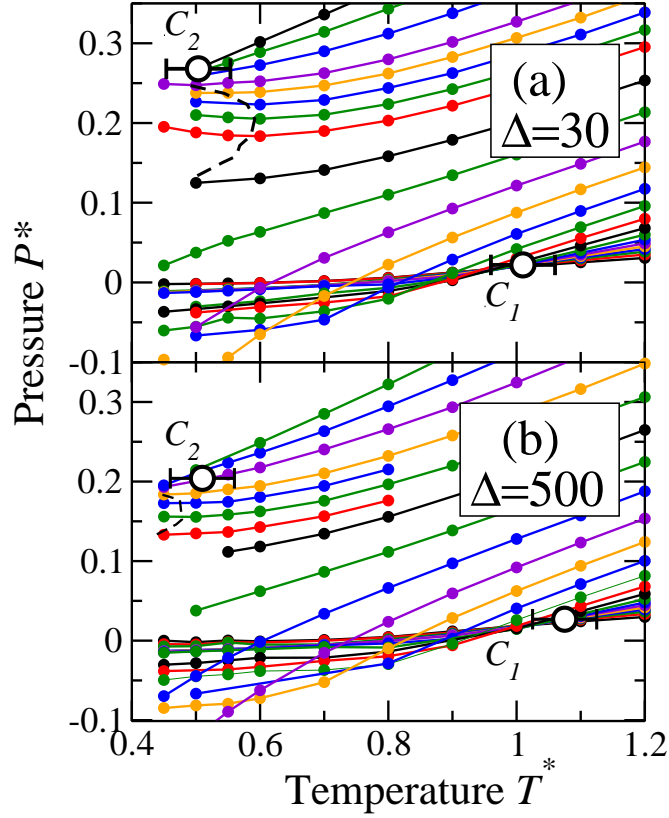}
\end{center}
		\caption{The $P^{*}-T^{*}$ phase diagram for
                  $\Delta=30$ (a) and $\Delta=500$
                  (b). In both panels, we show    
%                  At $T^{*}=1.3$
isochores (with constant density separation $\delta \rho^*=0.01$)
%correspond, from bottom to top to 
from
$\rho^{*}=0.04$ (at high $T^*$ and low $P^*$) to 
%                  $0.05$, $0.06$, $0.07$, $0.08$, $0.09$, $0.10$,
%                  $0.11$, $0.12$, $0.13$, $0.14$, $0.15$, $0.16$,
%                  $0.17$, $0.18$, $0.19$, 
$0.20$, from $0.205$
%, $0.21$,
%                  
to $0.215$ (with $\delta \rho^*=0.005$), and (with $\delta \rho^*=0.01$)
from 
$0.22$
%, $0.23$, $0.24$, 
to $0.25$ (at low $T^*$ and high $P^*$). The points
                  where isochores cross correspond to the gas-liquid and
                  LDL-HDL critical points ($C_1$ and $C_2$,
                  respectivelly). The black dashed line is a guide for
                  the eye estimating the TMD line, correspondig to the
                  line of minima along the isochores.} 
\label{f3}             
\end{figure}

For all the considered values of $\Delta$, we find a temperature of
minimum pressure along the isochores near the LDL-HDL critical point
$C_{2}$ (Fig.~\ref{f3}). These temperatures correspond to the TMD line
at constant $P$. We find that the anomalous region where $\rho$
decreases for decreasing $T$ shrinks for increasing $\Delta$ and
possibly tends to collapse onto one single point in the $P^{*}-T^{*}$
plane for $\Delta\rightarrow\infty$. This result would be consistent
with the behaviour of the DSW potential, that does not show density
anomaly \cite{Fr01,r34,r35,r36}.  

\subsection{Diffusion anomaly}

\begin{figure}%[ht]
\begin{center}\includegraphics[clip=true,scale=0.375]{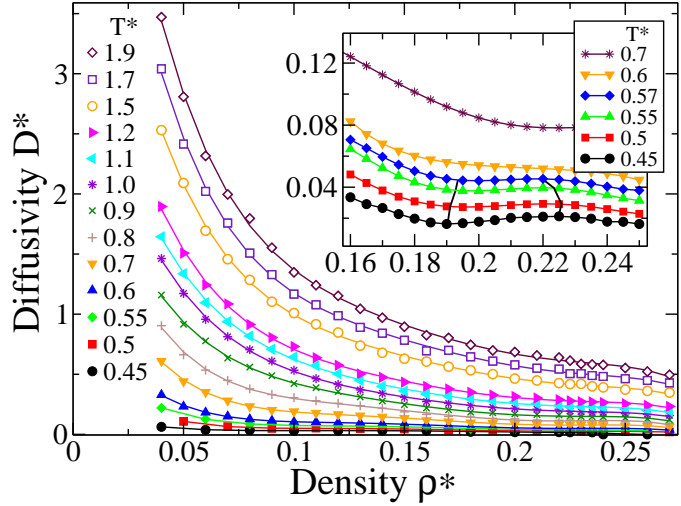}
\end{center}
\caption{Diffusion coefficient $D^{*}$ as a function of density
  $\rho^{*}$  for
  $\Delta=30$ and isotherms from
          $T^*=0.45$  (bottom) to $T^*=1.9$ (top).
Inset: Isotherms from $T^*=0.45$  (bottom) to $T^*=0.7$ (top)  
Black solid lines delimit the anomalous region where $D^{*}$ 
  grows from $D_{\rm min}$ to $D_{\rm max}$
for increasing density between
  $\rho^{*}_{D_{\rm min}}<\rho^{*}<\rho^{*}_{D_{\rm max}}$.} 
\label{f4}      
\end{figure}

For the family of potentials given by Eq.~(1) with the considered
values of $\Delta$ we find  an anomalous-diffusion region, i.e. a
region ($\rho_{Dmin}<\rho<\rho_{Dmax}$) where $D$ increases with
increasing density at constant $T$ (Fig.~\ref{f4}). 
The diffusion coeficient is calculated from the mean square
displacement of a single particle
\begin{equation}
 \langle\Delta r(t)^{2}\rangle\equiv\langle[r(t_{o}+t)-r(t_{o})]^{2}\rangle,
\end{equation}
where $t_o$ is any time at equilibrium and the average is over the
initial $t_o$ and over the particles in the system,
taking the long-time limit
\begin{equation}
 D=\lim_{t\rightarrow\infty}\frac{<\Delta r(t)^{2}>}{2dt}
\end{equation}
where $d=3$ is the dimension of the system.
We find that as in the case of the density anomaly, the region of
diffusion anomaly contracts as $\Delta$ increases. We find that the
diffusion anomaly region always encompasses the TMD line between
$\rho_{Dmin}$ and $\rho_{Dmax}$ (Fig.~\ref{f7}). 

\subsection{Structural anomaly}

\begin{figure}%[ht]
\begin{center}\includegraphics[clip=true,scale=0.375]{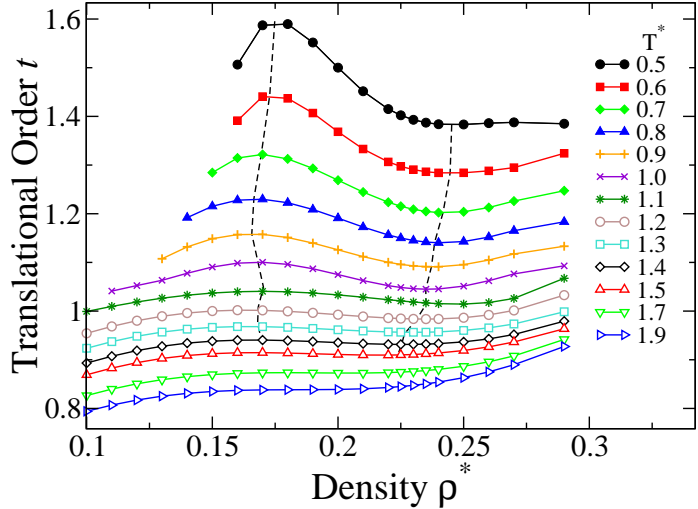}
\end{center}
	\caption{Translational order parameter $t$ as a function of
          density $\rho^{*}$ for $\Delta=30$ and isotherms from
          $T^*=0.5$ (top) to $T^*=1.9$ (bottom). Dotted lines joining the
          maxima and minima along the isotherms are a guide for the
          eye delimiting the region between $\rho_{t_{\rm max}}$ and
          $\rho_{t_{\rm min}}$, where $t$ decreases for increasing
          $\rho^{*}$.} 
\label{f5}
\end{figure}

The structural behaviour of the system is studied by calculating the
translational order parameter $t$ and the orientational order
parameter $Q_{6}$ defined in the following. 
The translational order parameter is defined as \cite{Sh02,Er01,Er03}
\begin{equation}
 t\equiv\int_{0}^{\infty}|g(\xi)-1|d\xi
\end{equation}
where $\xi\equiv r\rho^{1/3}$ is a reduced distance (in units of the
mean interparticle separation $\rho^{-1/3}$) and $g(\xi)$ is the
radial distribution function. Since in our simulations $g(\xi)\backsimeq1$ for
$\xi\geq\xi_{c}=8$, we integrate up to a cuttoff distance $\xi_{c}$.  
As the parameter $t$ depends only on the deviations of $g(\xi)$ from
unity, its value is sensible to long range periodicities. For an ideal
gas $g(\xi)$ is constant and equal to 1 and there is no translational
order ($t=0$). For a crystal phase $g(\xi)\neq1$ for long distances
and $t$ becomes large. 
While normal fluids show a monotonic increase of $t$ with increasing
density, for the CSW potential $t$ increases for $\rho<\rho_{t_{\rm max}}$,
and reaches a maximum at $\rho_{t_{\rm max}}$. Above $\rho_{t_{\rm
    max}}$, for
increasing $\rho$,  $t$ decreases until it reaches a minimum at
$\rho_{t_{\rm min}}$. For $\rho>\rho_{t_{\rm min}}$, $t$ recovers the normal
behaviour (Fig.~\ref{f5}). 

The local orientational order for a particle $i$ and an index $l$ is defined as \cite{St83}
\begin{equation}
 Q_{l}^{i}\equiv\left[\frac{4\pi}{2l+1}\sum_{m=-l}^{m=l}|(\bar{Y}^{i}_{lm})_{k}|^{2}\right]^{1/2}
\end{equation}
where $k$ is a fixed number of nearest neighbour particles and
$(\bar{Y}^{i}_{lm})_{k}\equiv\frac{1}{k}\sum_{j=1}^{k}Y_{lm}(r_{ij})$
is the average of the spherical harmonics $Y_{lm}$ with indices $l$
and $m$, evaluated over the
vectorial distance $r_{ij}$ between particles $i$ and $j$. 

The calculations involved in the evaluation of $Q_{l}^{i}$ can be
computationally expensive. In order to reduce the computational effort, we
express the spherical harmonics in terms of the Legendre Polynomials
$P_{l}$ obtaining 
\begin{equation}
 Q_{l}^{i}=\frac{1}{\sqrt{k}}\left[1+\frac{2}{k}\sum_{j=1}^{k-1}\sum_{h=j+1}^{k}
P_{l}(\cos\gamma_{jh}^{i})\right]^{1/2} ,
\end{equation}
where $\gamma_{jh}^{i}$ is the angle formed by the vectors
$\vec{r}_{ij}$ and $\vec{r}_{ih}$. 

A global
order parameter is obtained by averaging 
the local order $Q_{l}^{i}$ over all particles 
\begin{equation}
 Q_{l}\equiv\frac{1}{N}\sum_{i=1}^{N}Q_{l}^{i} ~.
\end{equation}
We study the case $l=6$ and $k=12$ as in Ref.~\cite{Ol06b}. 
For $l=6$ 
%and $k=12$ 
%we obtain
%\begin{equation}
%Q_{6}^{\rm s}=\frac{1}{\sqrt{12}}\left[1+\frac{1}{6}\sum_{i=1}^{11}\sum_{j=i+1}^{12}
%P_{6}(\cos\gamma_{ij}^{\rm s})\right]^{1/2}  ,
%\end{equation}
%where 
the sixth order Legendre Polynomial is given by
$P_{6}(x)=\frac{1}{16}\left(231x^{6}-315x^{4}+105x^{2}-5\right)$. 
For an
isotropic homogeneous system, $Q_{6}$ reaches its minimum value
$Q_{6}^{\rm ih}=1/\sqrt{k}=0.287$ ($k=12$), while for a fully ordered
f.c.c arrangement $Q_{6}^{\rm fcc}=0.574$.

\begin{figure}%[ht]
\begin{center}\includegraphics[clip=true,scale=0.375]{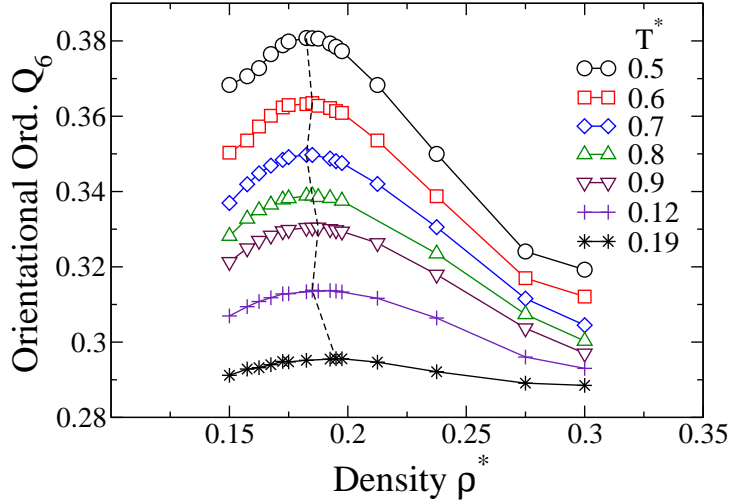}
\end{center}
	\caption{Orientational order parameter $Q_{6}$ as a function
          of $\rho^{*}$ for $\Delta=30$, for isotherms going from
          $T^*=0.5$ (top) to $T^*=1.9$ (bottom).
The dotted line is a guide for the eye joining the maxima along the isotherms.}
\label{f6}      
\end{figure}

The normal beahviour of $Q_{6}$ for a fluid without anomalies is to
increase monotonically when $\rho^{*}$ increases at constant $T^{*}$,
or when $T^{*}$ decreases at constant $\rho^{*}$. $Q_{6}$ decreases to
the value $Q_{6}^{\rm ih}$ at high $T^{*}$. For the CSW potential,
$Q_{6}$  has a non--monotonic behaviour along the isotherms with a
maximum at $\rho_{Q_{\rm max}}$ (Fig.~\ref{f6}). 

The density $\rho_{Q_{\rm max}}$ for each isotherm lies between
$\rho_{t_{\rm max}}$
and $\rho_{t_{\rm min}}$. In the area between $\rho_{Q_{\rm max}}$ and
$\rho_{t_{\rm min}}$
both order parameters decrease for increasing $\rho$, hence the liquid
gets more disordered with increasing density. This behaviour defines
the structural anomaly region ($\rho_{Q_{\rm
    max}}\leq\rho\leq\rho_{t_{\rm min}}$). 

By increasing the value of $\Delta$ we find that the region of
structural anomaly
does not contract, but tends assimptotically to a fixed region
in the $T^{*}-\rho^{*}$ plane (Fig.~\ref{f7}). 
This weak dependence on $\Delta$ suggests that the occurrence of
the structural anomaly does not disappear for very steep soft-core
potentials. 
This prediction  is consistent with the $s_2$ calculations 
that allow
de Oliveira et al. \cite{r15} to argue
that the
structural anomaly  should be observable also for 
the DSW potential, here considered as the limit of the CSW for $\Delta
\rightarrow\infty$.

\section{Discussion}

\begin{figure}%[ht]
\begin{center}\includegraphics[clip=true,scale=0.375]{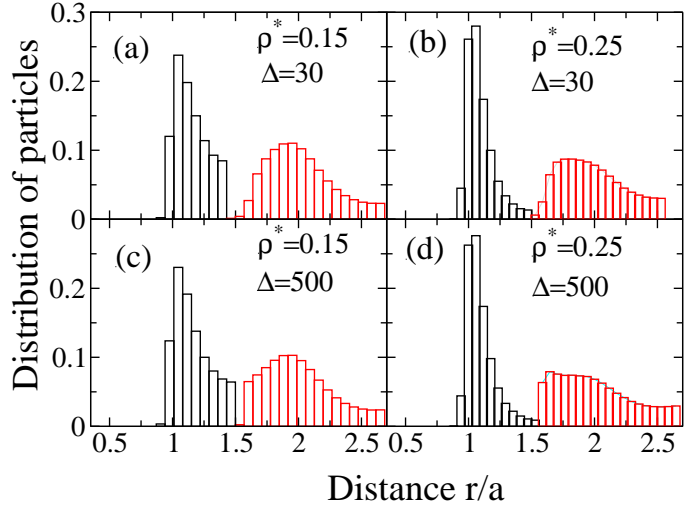}
\end{center}
\caption{Particle distribution for the first and second peaks of the
  $g(r)$, normalized by the area of each peak, at $T^{*}=0.5$: for
  $\Delta=30$ at (a) $\rho^{*}=0.15$ and (b) $\rho^{*}=0.25$, and for
  $\Delta=500$ at (c) $\rho^{*}=0.15$ and (d) $\rho^{*}=0.25$. The
  distance between the maxima of the first and the second peak
  decreases for increasing $\rho$ and for increasing $\Delta$.}
\label{f_histo}       
\end{figure}

To get more insight into the origin of the anomalies,
we study 
the relation between the changes of the
coordination shells of the radial distribution
function $g(r)$ and the appearence of the anomalies. 
First, we
perform a systematic study of the first two peaks of $g(r)$.
For example, for $\Delta=30$ we find that at densities within the LDL phase
(Fig.~\ref{f_histo}a)
the distribution of particles of the first peak is maximum at
$r_1/a\simeq1.05$, next to the hard-core. For the second peak, the
distribution is maximum at $r_2/a\simeq1.95$, at about the minimum
of the atractive well. At  densities within the HDL phase
 many particles overcome the shoulder approaching the hard-core. 
For $\rho^{*}=0.25$
(Fig.~\ref{f_histo}b) the maximum of the first peak increases without
changing its position, at $r_1/a\simeq1.05$, while the
distribution of the second peak is slightly flattened with a maximum
around $r/a\simeq 1.75$.
This change by increasing density 
shows that particles in the attractive well of the potential 
accumulate close to the soft-core distance, giving rise to an asymetric
distribution of particles around the minimum of the attractive well. 

For the highest $\Delta$ considered here, $\Delta=500$, we find a
similar result at the LDL density 
(Fig.~\ref{f_histo}c), but at the density $\rho^{*}=0.25$ (Fig.~\ref{f_histo}d), approaching
%the density of 
the LDL-HDL critical point,
the distribution of particles at the
second peak~\cite{note} displays a flat region in the range $1.7\leq
r/a\leq2.0$ and a maximum around the soft-core distance at
$R_R/a=1.6$. This is consistent with the fact that for $\Delta=500$ 
the potential has a more penetrable soft-core for 
$r/a>1.6$ with respect to the case $\Delta=30$. 
As a consequence, the
effective soft-core distance for the potential with $\Delta=500$ is
smaller than that for the potential with $\Delta=30$. 

This qualitative observation gives rise to a quantitative criterion for
the prediction of the occurrence of the density and diffusion anomaly
when we generalize what has been proposed by Yan et
al. for potentials with a repulsive ramp \cite{YanBul06}. 
%between distances 
They observe that, if $a$ and $b$ are respectively the starting and
ending distance of the repulsive ramp, 
 the anomalies of density and
diffusion occur only when $0.5<a/b<6/7=0.86$. This is consistent with the fact
that for real water in standard conditions the ratio between the
distances of the first two peaks of the oxygen-oxygen $g(r)$ is about
$0.6$, while in normal liquids it is about $0.5$ \cite{YanBul06}. 

A straightforward application of the the Yan's criterion to our case
is not obvious. One could assume that the relevant ratio in the CSW
potential is $a/R_A=0.5$, or as an alternative $R_R/R_A=0.8$. With both
choices one would conclude that Yan's criterion does not
hold for the CSW potential because it would always predict the
absence (with the first choice) or the occurrence (with the second
choice) of the anomalies independently on $\Delta$. 

Hence, to generalize the Yan's criterion to cases in which 
the effective soft-core distance changes with density or $\Delta$, as
in the present case, we
adopt as characteristic length-scales the maxima of the first two
peaks of the $g(r)$ within the HDL phase and define the 
ratio $\lambda(\Delta)\equiv
r_1(\rho)/r_2(\rho)|_{\rho_{\rm HDL}}$ at a temperature
below, or at about, the LDL-HDL critical point.
The ratio $\lambda$ is larger
for more penetrable potentials, i. e. for larger $\Delta$ in the case
of the CSW potential.

With this generalized definition, we conclude that the empirical
criterion proposed by Yan et al. \cite{YanBul06} is valid also 
for all the cases considered in our work. For the cases in
Fig.~\ref{f_histo} it is, for example,
$\lambda(\Delta=30)=0.55<\lambda(\Delta=500)\simeq 0.66$, where the
approximation symbol is used because for $\Delta=500$ we can evaluate
$\lambda$ only at densities approaching 
$\rho_{\rm HDL}(T)$.~\cite{note}
We observe that by increasing $\lambda$ the regions in the plane
$(T-\rho)$ where the density and diffusion anomalies occur reduce in 
a sensible way, while the region of structural anomalies
is almost unaffected by the change of $\lambda$ (Fig.~\ref{f7}).

The considerations above are possibly consistent with the idea that
the structure within the first two coordination shells is controlling
the occurrence of the anomalies. In the following we will show that
this concept is not leading to any general criterion for the occurrence of the
anomalies and that the anomalies are, instead, regulated by the
structure at a larger scale. 

To this goal, we test an idea 
%about  the relation between the change of the first two peaks of the $g(r)$
%second and higher coordination shells of the radial distribution
%function and the appearence of the anomalies  \cite{OlNetz08,YanBul06,KMGT08}.
%In particular, it has been 
proposed in Ref.~\cite{OlNetz08} suggesting 
that a necessary condition for the appearence of 
the
anomalies 
%appear only 
%when there 
is a transference of particles from the second shell of the
fluid to the first shell under isothermal compression. According to
this criterion
%if $r_{1}$ and $r_{2}$ are the positions of the first
%and second peak of the $g(r)$ respectivelly, then 
all the anomalies
appear for the range of $\rho$ and $T$ where  
\begin{equation}
\Pi_{1,2}\equiv  \left.\frac{\partial
     g(r)}{\partial\rho}\right|_{r_{1}}\times\left.\frac{\partial
     g(r)}{\partial\rho}\right|_{r_{2}}<0~,
\label{grcondition}
\end{equation}
i. e. Eq.(\ref{grcondition}) is a necessary condition.
On the other hand, if $\Pi_{1,2}\geq 0$, no
anomalies should be observed, i. e. the condition in
Eq.(\ref{grcondition}) is also sufficient.\cite{OlNetz08}

\begin{figure}%[ht]
\begin{center}\includegraphics[clip=true,scale=0.6]{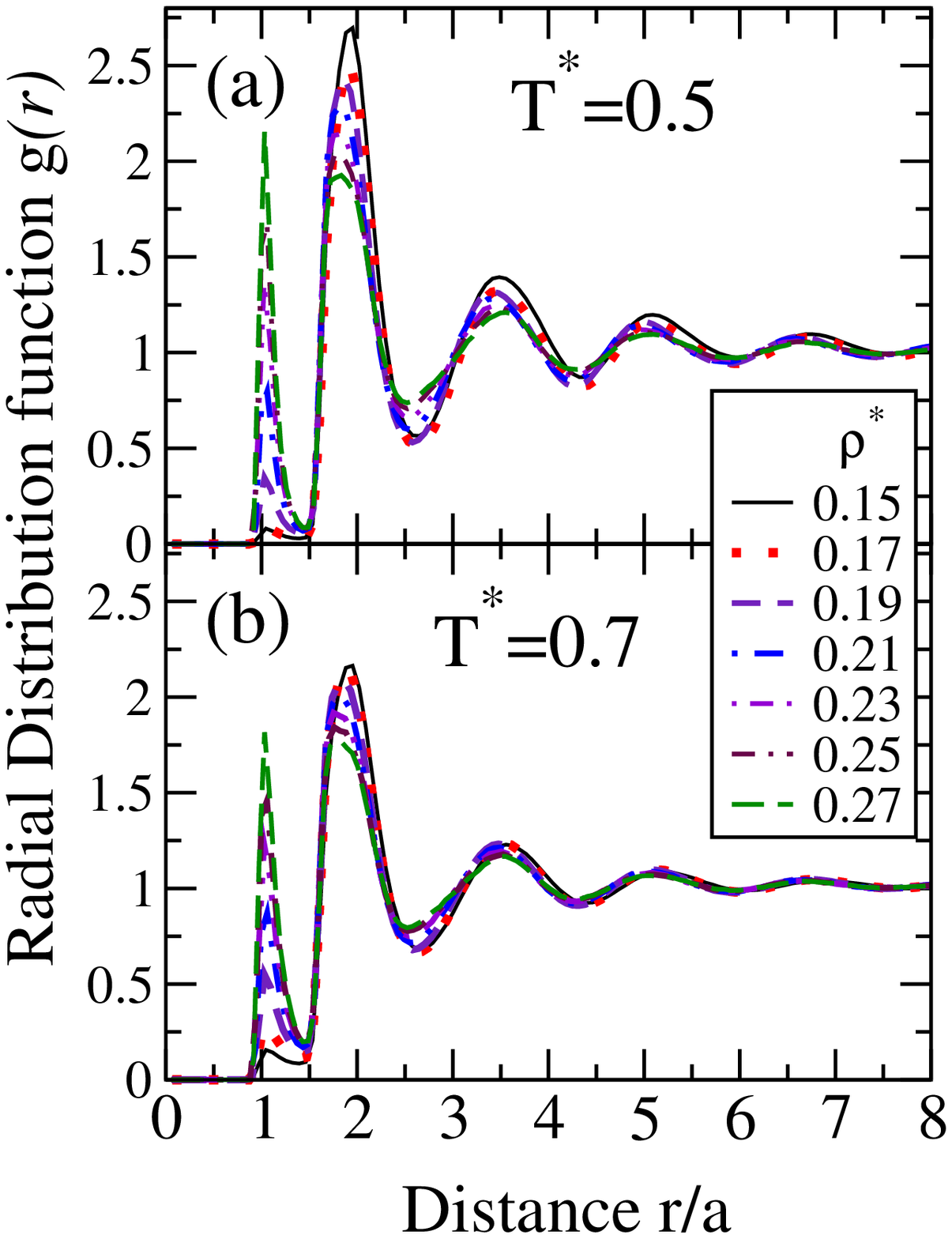}
\includegraphics[clip=true,scale=0.6]{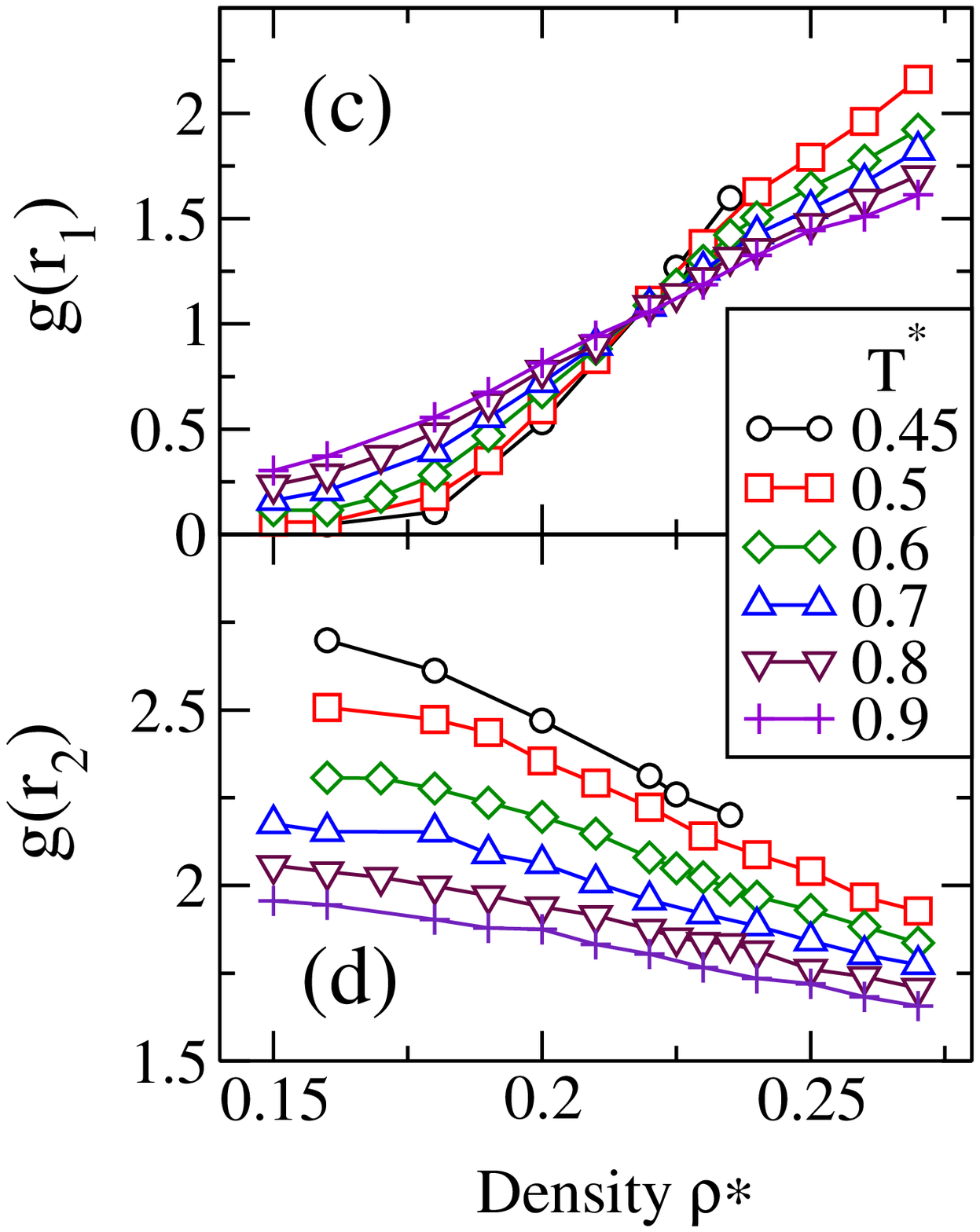}
\end{center}
		\caption{Change of the radial distribution function $g(r)$ for
                  $\Delta=30$ with $\rho^*$ and $T^*$. The
                  $g(r)$ at  densities $0.15\leq\rho^{*}\leq 0.27$ for (a)
$T^{*}=0.5$ and (b)
$T^{*}=0.7$. 
(c) The height of the first peak of $g(r)$, i. e. $g(r_1)$, always increases for
increasing $\rho^{*}$, but 
                  has a more gradual increase at higher $T$.
(d) The height of the second peak of $g(r)$, i. e. $g(r_2)$, always
decreases for increasing $\rho^{*}$ with a weak dependence on $T$.
}
\label{fgr_D30}       
\end{figure} 

\begin{figure}%[ht]
\begin{center}\includegraphics[clip=true,scale=0.65]{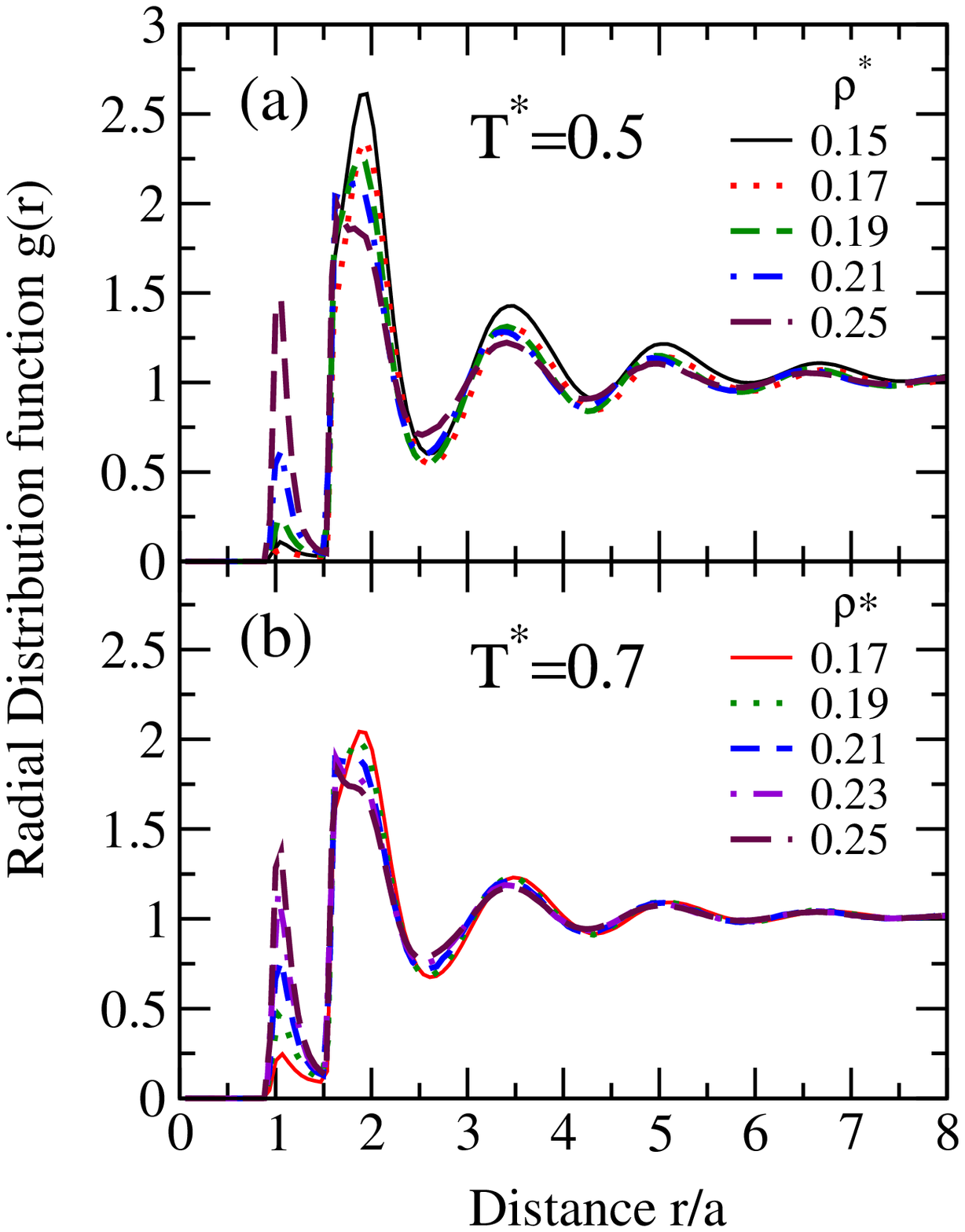}
\end{center}
	\caption{Radial distribution function $g(r)$ for
          $\Delta=500$. For $T^{*}=0.5$ (a) and for $T^{*}=0.7$ (b) the
          behaviour of $g(r)$ at different $\rho^{*}$ is comparable
          to the case in Fig.\ref{fgr_D30}, the main difference
          being the appearance of a fine structure in the first two
          peaks at high $\rho^{*}$, resembling the $g(r)$ of the DSW
          potential \cite{r34}.} 
\label{fgr_D500}      
\end{figure}

To study the implications of Eq.~(\ref{grcondition}), we analyse
the behaviour of the CSW $g(r)$ upon isotermal
compression (Fig.~\ref{fgr_D30} and Fig.~\ref{fgr_D500}). For all the
considered values of $\Delta$ and $\rho^{*}$ we find that it is always
$\Pi_{1,2}<0$, even for $T$ and $\rho$ outside the regions
of the anomalies. 
Therefore, the condition in
Eq.~(\ref{grcondition}) could be necessary, but not sufficient,
for the appearance of 
the anomalies. This observation is consistent with the result for the
DSW potential, where Eq.~(\ref{grcondition}) is satisfied (see Fig.~17
in Ref.~\cite{r34}) but no
density anomaly is observed \cite{Fr01,r34,r35,r36}. 

\begin{figure}%[!htp]
\begin{center}
\includegraphics[clip=true,scale=0.6]{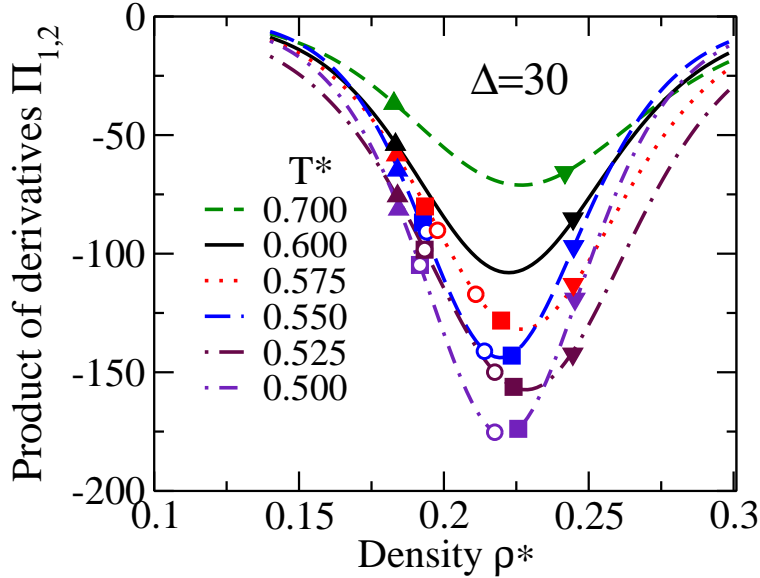}
\end{center}
	\caption{Product $\Pi_{1,2}$ in Eq.(\ref{grcondition})
%of the derivatives of the 
%first  and second peak of the $g(r)$
as a function of 
          density for $\Delta=30$ and different temperatures.
%and (b) $T^*=0.5$ and different  values of  $\Delta$.
Symbols mark the lowest and the highest density 
where at each $T$ a specific anomaly is observed: 
$\blacktriangle$ for the maximum $Q_6$ and $\blacktriangledown$ for
the minimum $t$, marking the structural anomaly region,
$\blacksquare$ for the limits of the diffusion anomaly region,
$\ocircle$ for the limits of the density anomaly region.
For other values of $\Delta$, not shown here, we find the same
qualitative behavior.}
\label{30}
 \end{figure}

To understand better why Eq.~(\ref{grcondition}) is not a sufficient
condition for the anomalies, we 
study how $\Pi_{1,2}$
changes with $\rho$ and $T$ (Fig.~\ref{30}). 
For each $T$, we find that $\Pi_{1,2}(\rho)$ is not monotonic, 
consistent with the fact that $\Pi_{1,2}(\rho)$ is expected to be  
negative around the anomaly regions and to be positive at smaller
and larger $\rho$. 
We observe that $\Pi_{1,2}(\rho)$ reaches a lower minimum 
for lower $T$, i. e.  $\Pi_{1,2}$ is more negative where the anomalies
are stronger. However, away from its minimum $\Pi_{1,2}(T)$ is not 
a monotonic function of $T$ at constant $\rho$.
Moreover, we
find that $\Pi_{1,2}(\rho)$ is not symmetric
with respect to the center of the anomalous regions.
Instead, $\Pi_{1,2}$ reaches its minimum close to the maximum density
with diffusion anomaly (Fig.~\ref{30}). 
We also observe that for $\Pi_{1,2}$ there is no clear threshold 
below which an anomaly appears, contrary to the case of $\Sigma_2$
derived from the excess entropy.  
%For example, 
%for $T=0.575$ the diffusion anomaly occurs for $\Pi_{1,2}<-81$ at low
%density, but not at high density and not for $T=0.6$.
All these observations suggest that there is 
no condition on $\Pi_{1,2}$ that is sufficient for the
occurrence of the anomalies. More generally, these results suggest
also that the onset of the anomalies is related not only to the
behavior of the first and second coordination shell of the $g(r)$, but
also to the
behavior of the higher coordination shells, as recently observed by
Krekelberg et al. for 
other models of anomalous liquids, including water \cite{KMGT08}.

\begin {figure}
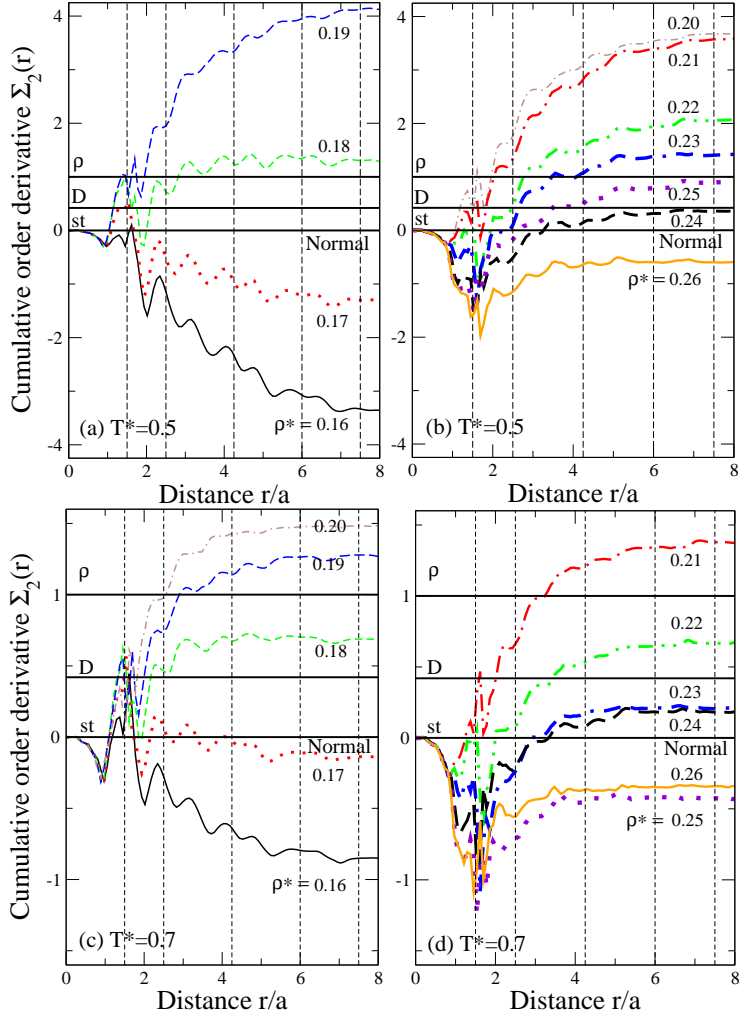
%[!htp]
\begin{center}
\includegraphics[clip=true,scale=0.4]{epsilon_r_T05_D30.eps}
\includegraphics[clip=true,scale=0.4]{epsilon_r_T07_D30.eps}
\end{center}
\caption{The cumulative order derivative $\Sigma_2(r)$ calculated for
  the case $\Delta=30$, for different temperatures and densities. 
(a) For  $T^{*}=0.5$ and density going from 
$\rho^{*}=0.16$ (lower line)
to  
$\rho^{*}=0.19$ (upper line).
(b) For  $T^{*}=0.5$ and density going from 
$\rho^{*}=0.20$ (upper line)
to  
$\rho^{*}=0.26$ (lower line).
(c) For  $T^{*}=0.7$ and density going from 
$\rho^{*}=0.16$ (lower line)
to  
$\rho^{*}=0.20$ (upper line).
(b) For  $T^{*}=0.7$ and density going from 
$\rho^{*}=0.21$ (upper line)
to  
$\rho^{*}=0.26$ (second lower line).
The values of $\rho^*$ are indicated by the labels near the 
calculated lines.
Horizontal continuous lines with labels {\it st}, $D$ and $\rho$ mark
the thresholds for structural anomaly,  diffusion anomaly, and density
anomaly, respectively.
Vertical dashed lines mark the coordination shells as defined by the
minima of $g(r)$ in Fig.~\ref{fgr_D30}a,b.} 
\label{Int-Sigma2}
   \end {figure}

\begin {figure}
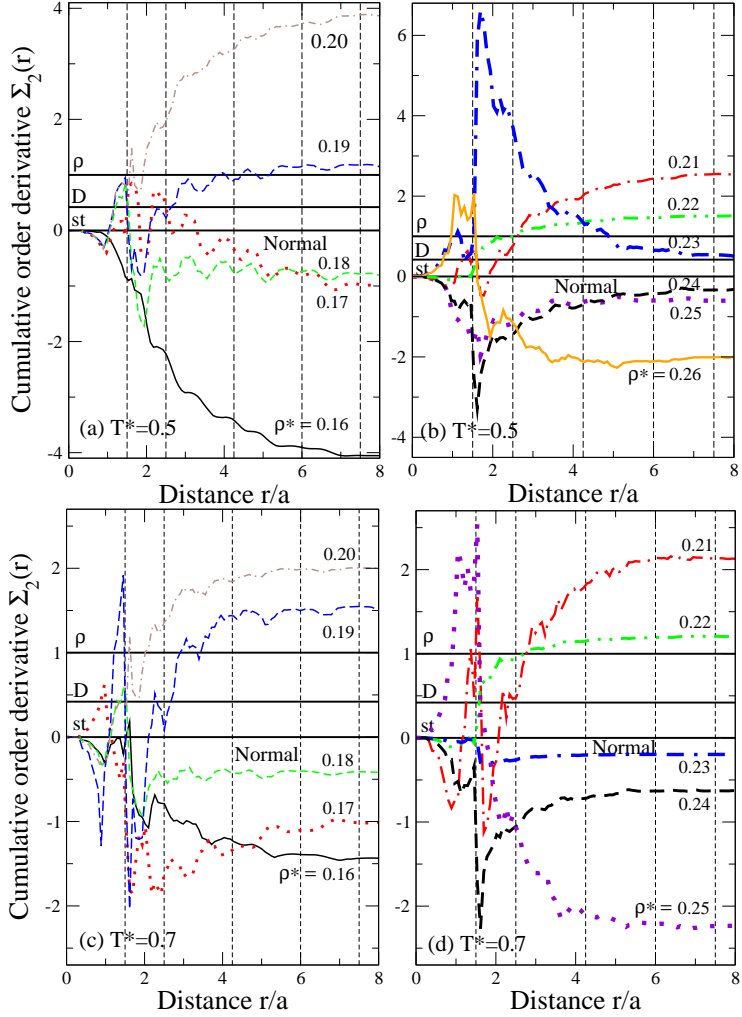
%[!htp]
\begin{center}
\includegraphics[clip=true,scale=0.4]{epsilon_r_T05_D500.eps}
\includegraphics[clip=true,scale=0.4]{epsilon_r_T07_D500.eps}
\end{center}
\caption{As in Fig.~\ref{Int-Sigma2}, but for  $\Delta=500$.
(a) For  $T^{*}=0.5$ and density going from 
$\rho^{*}=0.16$ (lower line)
to  
$\rho^{*}=0.20$ (upper line).
(b) For  $T^{*}=0.5$ and density going from 
$\rho^{*}=0.21$ (upper line)
to  
$\rho^{*}=0.26$ (lower line).
(c) For  $T^{*}=0.7$ and density going from 
$\rho^{*}=0.16$ (lower line)
to  
$\rho^{*}=0.20$ (upper line).
(b) For  $T^{*}=0.7$ and density going from 
$\rho^{*}=0.21$ (upper line)
to  
$\rho^{*}=0.25$ (lower line).
Vertical dashed lines mark the coordination shells as defined by the
minima of $g(r)$ in Fig.~\ref{fgr_D500}a,b.} 
\label{Int-Sigma2-D500}
\end{figure}

To better understand the relevance of the outer shells, following
Ref.~\cite{KMGT08} we perform an analysis of the excess entropy shell
by shell, but instead of calculating only the cumulative order
integral, as in Ref.~\cite{KMGT08},
\begin{equation}
 s_{2}(r)\equiv-2\pi\rho\int_0^r[g(r')\ln g(r')-g(r')+1]r'^{2}dr'
\label{s2r}
\end{equation}
here we consider also its derivative
\begin{equation}
\Sigma_2(r)\equiv \left.\frac{\partial s_{2}(r)}{\partial \ln\rho}\right|_{T}~,
\label{Sigma2r}
\end{equation}
where $s_{2}(r)\rightarrow s_2$ and  $\Sigma_2(r)\rightarrow \Sigma_2$
for $r\rightarrow \infty$. 
Our analysis 
shows that $\Sigma_2(r)$ presents
many oscillations within each coordination shell
(Fig.~\ref{Int-Sigma2}, Fig.~\ref{Int-Sigma2-D500}).  In several cases
different shells contribute with a different sign to the total
$\Sigma_2$ and in some cases the relevant contribution to the
large--$r$ value of $\Sigma_2(r)$ comes from the third or higher
coordinations shell.

For example, in the case with $\Delta=30$  we observe that
at low density (Fig.~\ref{Int-Sigma2}a,c for $T^*=0.5$ and $0.7$,
respectively) the major 
contribution to the structural anomaly comes from the first and the
second shell, although for $\rho^*=0.17$ at  $T^*=0.7$
(Fig.~\ref{Int-Sigma2}c) it is only after the forth shell that 
$\Sigma_2(r)$ stops to oscillate around the threshold for the
structural anolmaly. On the other hand, the relevance of the third and higher
coordination shells is very evident at higher density
(Fig.~\ref{Int-Sigma2}b,d), where the sign of $\Sigma_2(r)$ changes in
some cases within the third shell ($\rho^*=0.24$ at $T^*=0.5$,
$\rho^*=0.23$ and $\rho^*=24$ at $T^*=0.7$).

Also for the diffusion and the density anomaly, the relevance of the
outer shells is evident, especially at high $\rho$. For example,
in Fig.~\ref{Int-Sigma2}b there are several cases where the thresholds
for  diffusion and density anomaly are reached only within the third shell.
While in the majority of cases $\Sigma_2(r)$ saturates within the
first fifth shells, we observe also cases where the sixth and
possibly higher coordination shells are relevant for the final value of $\Sigma_2$,
e. g. for $\rho^*=0.24$ and $\rho^*=0.25$ at  $T^*=0.5$
(Fig.~\ref{Int-Sigma2}b) with 
respect to the empirical thresholds for the density anomaly and the
diffusion anomaly, respectively.
Therefore, in general is not possible to establish the occurrence
of the anomalies on the base of the local structure (first and second
shell), but it is necessary the structural information at much
larger scale (forth and outer shells).

From Fig.~\ref{Int-Sigma2}c,d 
we observe that the empirical
thresholds for diffusion anomaly and density anomaly fail to 
predict correctly the absence of these anomalies at
$T^*=0.7$. The comparison with Fig.~\ref{f7}b, indeed,  shows that
these anolmalies are not found at $T^*=0.7$, while the calculation of
$\Sigma_2(r)$, over more than two 
coordination shells, would predict the occurrence of
the diffusion anomaly for $0.17<\rho^*<0.23$ and
of the density anomaly for $0.18<\rho^*<0.22$
(Fig.~\ref{Int-Sigma2}c,d).
Nevertheless, note that the scale in Fig.~\ref{Int-Sigma2}c,d is 40\%
smaller than in Fig.~\ref{Int-Sigma2}a,b and $\Sigma_2(r)$
never exceeds $1.5$.

The same analysis for $\Delta=500$ 
presents even more striking oscillations of  $\Sigma_2(r)$ within the
first two coordination shells (Fig.~\ref{Int-Sigma2-D500}). This is,
for example, evident for 
for $\rho^*=0.19$ and $T^*=0.7$ (Fig.~\ref{Int-Sigma2-D500}c)
at $r/a=1.5$,
where the sudden change of slope of the penetrable soft-core gives
rise to a large negative contribution to $\Sigma_2(r)$. However, the
contribution of the outer shells leads to positive $\Sigma_2$.
At the same $T^*$ and higher $\rho^*$, similar discontinuities at 
$r/a=1.5$ lead to negative $\Sigma_2$, as for $\rho^*=0.25$
(Fig.~\ref{Int-Sigma2-D500}d).
A remarkably large positive contribution to $\Sigma_2$ at $r/a=1.5$ is
observed at $T^*=0.5$ and $\rho^*=0.23$ (Fig.~\ref{Int-Sigma2-D500}b),
in correspondence to the limit of the region of structural anomaly
(Fig.~\ref{f7}d). However, also in this case the contribution coming
from the outer shells is comparable to that coming from the second
shell. 

For $\Delta=500$ the system displays structural anomaly at any $T$,
but diffusion and density anomalies are present only below $T^*=0.5$ 
(Fig.~\ref{f7}d). Nevertheless, from Fig.~\ref{Int-Sigma2-D500} we observe that
the empirical thresholds of $\Sigma_2$ for diffusion
and density anomalies fail at any $T$, as seen for $\Delta=30$ 
(Fig.~\ref{Int-Sigma2}) at higher $T$ only. Also for $\Delta=500$ the
higher the $T$ the lower the saturation value of $\Sigma_2(r)$, as for
$\Delta=30$.

Finally, we observe that in all the cases we have analyzed, when
$\Sigma_2(r)$ saturates to a value lower than $2.5$, both diffusion
and density anomalies are not observed. The only possible exception is
observed 
for $\Delta=500$ at $T^*=0.5$ and $\rho^*=0.20$
(Fig.~\ref{Int-Sigma2-D500}a), where $\Sigma_2=4$. However, this
state point is at the border of the region where both  diffusion
and density are anomalous (Fig.~\ref{f7}d).

\section{Summary and Conclusions}

We perform a systematic analysis of the soft-core 
CSW potential~\cite{franzCSW1} in Eq.~(1) by varying the slope $\Delta$
of the repulsive shoulder. We consider values from $\Delta=15$
(smooth and less penetrable at large distance)~\cite{r15} to $\Delta=500$ 
(almost discontinuous and more penetrable at large distance),
we calculate their phase diagrams and the occurrence of the anomalies,
and we
compare  with the results for the discountinuous (DSW) version of
the potential~\cite{Fr01}.

All the phase diagrams for different values
of $\Delta$ display at low density
the gas-liquid first order phase transition, 
ending in a critical point $C_1$, 
and at higher density
the LDL-HDL first order phase transition,
ending in a critical point $C_2$.
By increasing
$\Delta$, both critical points aproach the values of the two critical points for 
the DSW potential (Table I). In this sense,
the DSW can be considered as a limiting
case of the CSW potential. 

For any $\Delta$ we observe the occurrence of the density anomaly,
diffusion anomaly, and structural anomaly with the same hierarchy as
for water \cite{An76,Er01}, i. e. the density anomaly occurrs in the inner
region and the structural anomaly in the most external region.
The difusion anomaly region always includes the density anomaly
region, being the two regions
only slightly separated, with decreasing
separation for decreasing $\rho$ and increasing $\Delta$ (Fig.~\ref{f7}). 

The region of structural anomaly is always
present and almost independent of the value of $\Delta$. It slightly
contracts toward a limiting finite region for increasing $\Delta$. 
On the other hand, the regions of diffusion and density anomalies
narrow for increasing $\Delta$.
A similar behavior has been found recently for
purely repulsive soft-core potentials \cite{EPLOliveira09}.
Our finding gives a rationale for the absence of density anomaly
for the DSW potential \cite{Fr01,r34,r35,r36}, here seen as the limit
of the CSW for $\Delta\rightarrow \infty$.  
%
%For all values of $\Delta$ the LDL-HDL critical point remains, 
%as well as for the DSW potential, where it has been shown that there
%is no density anomaly \cite{Fr01}, 
%consistent with the observation
%that the occurrence of the anomaly in density is not necessary for
%the presence of the LDL-HDL critical point \cite{Fr01}. 

%Therefore, the
%anomalies depend on the slope of the shoulder, i.e. on its sofftness
%and the hierarchy of anomalies is always the same as for water.  

We find that the excess entropy analysis gives an approximate estimate
of the anomalous regions.
However, comparison with Ref.~\cite{r15}
shows that the behaviour of $\Sigma_{2}$ for the DSW potential is
still far from what we obtain for $\Delta=500$, consistent with a slow
convergence of the CSW toward the DSW potential when $\Delta$ is
increased.

We tested some ideas recently suggested as possible explanations for the
anomalous behavior of other soft-core potentials. 
First, we generalized to our CSW potential the 
Yan et al. \cite{YanBul06} criterion for the occurence of the
diffusion and density
anomalies in ramp-like potentials with two characteristic
length-scales. To this goal we propose to use as effective length-scales the 
distances of maxima of the 
first and second peak of $g(r)$ within the HDL phase. While Yan et
al. criterion does not apply to our CSW potentials, 
our generalized criterion 
is consistent with our data for different values of $\Delta$ and is
possibly valid for other continuous potentials with competing length-scales.

Next,
we observe that the condition in Eq.~(\ref{grcondition}) proposed by
de Oliveira et al. \cite{OlNetz08} is not enough to mark the occurence of the
anomalies. This condition relies on the structural changes of the first and
second coordination shell, while from our calculations we conclude that the
occurrence of anomalies is related to long range structural
changes. To show this result we consider the 
derivative $\Sigma_2(r)$, Eq.~(\ref{Sigma2r}), of the 
cumulative order
integral introduced by Krekelberg et al. \cite{KMGT08}. The study of
$\Sigma_2(r)$ shows that the anomalous behavior of the potential is
dominated at high $\rho$ and low $T$ by the structural changes up to
the fourth coordination shell, and possibly by even outer shells. We
observe that by increasing 
$\Delta$ (more penetrable and sharp soft-core) the contribution to  $\Sigma_2$
from the first and second shell becomes more relevant. 
We conclude that the gradual disappearing of the anomalies for large
$\Delta$ is regulated by the appearence of a more structured
short-range order (first and second coordination shell) that dominates
over the effect of the larger-range structure. 

\subsection*{Acknowledgements}
We thank for financial support 
the Spanish Ministerio de Ciencia e Innovaci\'on
Grants No. FIS2009-10210 (co-financed FEDER).

\bibliographystyle{aip}
%\bibliography{bibliografiapol}

\begin{thebibliography}{10}
\bibitem{note}
For $\Delta=500$ we cannot analyze data at density higher than
$\rho^{*}=0.25$ for the occurrence of spontaneous crystal 
nucleation. At density $\rho^{*}=0.25$ we observe the coexistence of 
the LDL phase and, in smaller ammount, the HDL phase. As a
consequence the $g(r)$ averaged over the entire system reflects the
presence of the HDL phase.
\end{thebibliography}

\end{document}